\documentclass[a4paper]{article}
\usepackage{graphicx,amsmath}
\usepackage{txfonts}
\usepackage[sort]{natbib}
\bibpunct{(}{)}{;}{a}{}{,} 
\usepackage[margin=2.5cm]{geometry}

\def\araa{ARA\&A}%
\def\apj{ApJ}%
\def\apjl{ApJ}%
\def\aap{A\&A}%
\def\mnras{MNRAS}%
  \renewenvironment{thebibliography}[1]{%
    \begin{oldthebibliography}{#1}%
      \setlength{\parskip}{0ex}%
      \setlength{\itemsep}{0ex}%
  }%
  {%
    \end{oldthebibliography}%
  }

\begin{document}
   \author{H.~M.~Cuppen$^{1}$, E.~F.~van Dishoeck$^{1,2}$, E.~Herbst$^{3}$, and A.~G.~G.~M.~Tielens$^{1}$\\
   \normalsize
                $^{1}$Leiden Observatory, Leiden University, PO Box 9513, 2300 RA Leiden, The Netherlands\\
   \normalsize$^{2}$Max-Planck-Institut f\"ur Extraterrestrische Physik, Giessenbachstrasse 1, 85748 Garching, Germany\\
   \normalsize$^{3}$Departments of Physics, Astronomy, and Chemistry, The Ohio State University, Columbus, OH 43210, USA}
\title{Microscopic simulation of methanol and formaldehyde ice formation in cold dense cores}

\date{}
   \maketitle





  \begin{abstract}
    Methanol and its precursor formaldehyde are among the most studied organic molecules in the interstellar medium and are abundant in the gaseous and solid phases. We recently developed a model to simulate CO hydrogenation via H atoms on interstellar ice surfaces, the most important interstellar route to H$_2$CO and CH$_3$OH, under laboratory conditions. 
   We extend this model to simulate the formation of both organic species under interstellar conditions, including  freeze-out from the gas and hydrogenation on surfaces.  Our aim is to compare calculated abundance ratios with observed values and with the results of prior models.  
    Our model utilises the continuous-time, random-walk Monte Carlo method, which --- unlike other approaches --- is able to simulate microscopic grain-surface chemistry over the long timescales in interstellar space, including the layering of ices during freeze-out. 
     Simulations under different conditions, including density and temperature, have been performed.  We find that H$_2$CO and CH$_3$OH form efficiently in cold dense cores or the cold outer envelopes of young stellar objects. The grain mantle is found to have a layered structure with CH$_3$OH on top. The species CO and H$_2$CO are found to exist predominantly  in the lower layers of ice mantles where they are not available for hydrogenation at late times. This finding is in contrast with previous gas-grain models, which do not take into account the layering of the ice.  Some of our results can be reproduced by a simple quasi-steady-state analytical model that focuses on the outer layer.
Observational solid H$_2$CO/CH$_3$OH and CO/CH$_3$OH abundance ratios in the outer envelopes of an assortment of young stellar objects agree reasonably well with our model results, which also suggest that the large range in CH$_3$OH/H$_2$O observed abundance ratios is due to variations in the evolutionary stages. Finally, we conclude that the limited chemical network used here for surface reactions apparently does not alter the overall conclusions.
\end{abstract}


\section{Introduction}

Methanol and its precursor formaldehyde are among the most studied organic molecules in the interstellar medium (ISM). Both species have been observed in the gas phase and as constituents of ice mantles. Whereas  gas-phase detections of both molecules have been numerous, \citep[\emph{e.g.},][]{ vanderTak:2000,Maret:2004,Schoier:2004,Jorgensen:2005II,Maret:2005} and methanol is a well-known constituent of the ice \citep{Allamandola:1992, Dartois:1999, Herbst:2009}, there are only a few secure solid-state detections of H$_2$CO.  These detections occur mostly in high-mass young stellar sources such as W33A, GL2136, and GL 989, and often only upper limits are given \citep{Gibb:2004,Boogert:2008}. For example, \cite{Pontoppidan:2004} found an upper limit for the H$_2$CO ice abundance in the outer envelope of the low-mass young stellar source \protect{Serpens SMM~4} of 5\% with respect to water ice. CH$_3$OH ice has a much higher abundance of 28\% with respect to water in this source, resulting in an H$_2$CO/CH$_3$OH ice ratio of less than 0.18. 
Typically, in sources where both species have been detected, the observed H$_2$CO/CH$_3$OH ratio is still smaller than unity,  with values ranging from 0.09 to 0.5.  Sub-millimetre observations of the gas in seven massive hot-cores show a similar trend: \cite{Bisschop:2007III} found a constant ratio of gas phase H$_2$CO/CH$_3$OH  between 0.13 and 0.28. 

 In this paper, we aim to develop a model that explains these
  observed abundances and abundance ratios and to study the
  implications for the conditions in regions where they are
  found. Following \cite{Tielens:1982} and \cite{Charnley:1992} and
  supported by numerous laboratory studies on hydrogenation of CO
  \citep{Hiraoka:2002,Watanabe:2002, Fuchs:2009}, we assume that
  methanol and formaldehyde are formed from hydrogenation of CO on
  grain surfaces. Observations have revealed that CO is present in
  three distinct components in interstellar ices: non-polar (pure) CO,
  CO mixed with CO$_2$ and CO as a trace in polar ice
  \citep{Tielens:1991, Chiar:1998, Pontoppidan:2008}. Here, we
  exclusively focus on methanol formation during the
  condensation/formation of the non-polar CO component. This component
  is thought to be formed when the cloud density increases and the
  gaseous atomic H/CO ratio decreases \citep{Tielens:1982}, and thus
  occurs at a later stage or deeper in the cloud than the water ice
  formation. This is consistent with the higher observed extinction
  threshold for CO ice formation compared with that of H$_2$O \citep{Whittet:2001}.
In the highest density regions ($>$ few $\times 10^5$ cm$^{-3}$), the
timescales for collisions of CO with the grains become so short that
most of the gaseous CO is removed from the gas. This so-called
`catastrophic' CO freeze-out has been observed directly through
infrared ice-mapping observations \citep{Pontoppidan:2006} and
indirectly through the lack of gas-phase CO (and other molecules) from
the densest parts of the core, \emph{e.g.}, L1544, B68
\citep{Caselli:1999,Bergin:2002}, as well as the accompanying rise in
H$_2$D$^+$ and the deuterated molecules. In such high density regions, the
bulk of the CO ice is in the pure form.
For this pure CO-ice component, we can thus be assured that CO is the main
reaction partner of accreted hydrogen and hence all the relevant
reactions have been studied experimentally. For the other CO ice
environments, competition with other reactions clearly plays an
important role, in particular hydrogenation of atomic oxygen to water.
We briefly comment on this competition in \S 5 but leave the formation
of methanol and formaldehyde in these other interstellar ice
environments to a future paper.

The formation route of methanol ice by hydrogenation of solid CO studied in
the laboratory consists of the following steps:
\begin{equation}
\label{methanol}
{\rm CO \rightarrow HCO \rightarrow H_{2}CO \rightarrow H_{2}COH \rightarrow CH_{3}OH},
\end{equation}
where the reactions involving CO and H$_{2}$CO possess small
activation energy barriers.  The experiments all show that for a range
of temperatures and H-fluences (time $\times$ flux) that are
comparable with fluences in the ISM, the H$_2$CO rate of production is
equal to or higher than the formation rate of CH$_3$OH. This result
appears to be in contradiction with the observational evidence. In a
recent paper {\citep{Fuchs:2009}}, we showed by means of microscopic Monte Carlo
simulations that high H-atom laboratory fluxes and low H-atom
interstellar fluxes do not result in the same production rate of
H$_2$CO and CH$_3$OH, even if the overall fluences are the same.  The
difference arises because most of the hydrogen atoms in the high-flux
regime form molecular hydrogen whereas in the interstellar regime the
hydrogenation reactions dominate \citep{Fuchs:2009}. The crossover
from H$_2$CO-rich conditions to CH$_3$OH-rich conditions occurs
therefore at much earlier times than expected from a direct
translation of the laboratory environment. Moreover, the simulations
presented in the \citet{Fuchs:2009} paper are for hydrogenation on
pure CO ice, a situation which is unlikely in the ISM.  Indeed, from
the final interstellar yield of only 4 monolayers (ML) obtained after
$ 2 \times 10^5$ yr, it was concluded that conversion of CO to
CH$_3$OH ice must occur simultaneously with the freeze-out and
building up of the CO layer, since hydrogenation after freeze-out does
not result in the high methanol abundances that are observed within
reasonable times. Furthermore, although hydrogen resides in dense
interstellar clouds predominantly in the form of H$_2$, a fraction is
present in atomic form together with gas phase CO, which suggests
simultaneous hydrogenation and freeze-out. The present paper presents
simulations of methanol formation under cold core conditions, where
hydrogenation and freezing out of the CO both occur.  These conditions
also pertain to the outer envelopes of YSO's.  Different temperatures
and initial abundances are used and the results are compared with
observational data on the H$_2$CO/CH$_3$OH gas and ice abundance
ratios. The laboratory results will further be extrapolated to lower
temperatures, which are more relevant to cold dense cores.

The models presented in this paper differ from previous models on the
formation of interstellar methanol ice in that they include details of
the surface structure of CO and utilise parameters obtained to
reproduce experiments in the laboratory setting {\citep{Fuchs:2009}}. In this manner, a set
of energy barriers for the different processes on the surface --
diffusion, desorption and reaction -- was obtained, which can now be
used in an interstellar environment.  Unlike the master equation
\citep{Stantcheva:2002}, rate equation \citep{Ruffle:2000}, and
macroscopic Monte Carlo \citep{Ruffle:2000} studies of methanol
formation, the continuous-time, random-walk (CTRW) Monte Carlo
technique accounts for the layering of the CO. This layering is
crucial, because the constant addition of fresh CO tends to protect
the underlying layers from hydrogenation and limits the time available
for reaction. Moreover, our current approach is different from
previous CTRW-Monte Carlo simulations on this system
\citep{Chang:2007}, both because we use newly determined barriers, and
because we account for the actual crystal structure of CO ice, instead
of utilising a simple cubic structure. On the other hand, the previous
study of \citet{Chang:2007} coupled the gas-phase chemistry with
grain-surface chemistry and used a larger grain surface reaction
network.

The paper is organised in the following way. The next section
describes the Monte Carlo method and introduces the input parameters
of the model and their origin. Section~\ref{results} presents the
model results for interstellar conditions covering the same
temperature regime as in the experiments (12.0-16.5~K) {\citep{Fuchs:2009}}. The focus of
the discussion is on how the abundances depend on various parameters,
especially the gas-to-dust ratio, and on the ice structure and
layering. Section~\ref{cold} extends the model to lower temperatures
of 8 and 10~K, which are more appropriate for cold dense cores. In
Section \ref{comparison} the model results are compared to a rate
equation model using similar input parameters and conditions, an
analytical steady-state model, other prior grain surface models, and
observations. Section \ref{comp reactions} discusses the effect of the
limited chemical network on the results by introducing more reactions.

\section{Monte Carlo method}   
The CTRW method has been described previously \citep[see, \emph{e.g.},][]{Chang:2005}.
A detailed overview of the Monte Carlo program and its input parameters can be found in \cite{Fuchs:2009}.   In brief, the technique simulates a sequence of processes that can occur on a grain surface, which is modelled as a lattice with the number of lattice sites determined by the size of the grain and the site density for the adsorbate CO. The order of this sequence is determined by means of a random number generator in combination with the rates for the different processes. These processes include deposition onto the surface, hopping from one lattice site to a nearest neighbour, desorption of the surface species,  and reactions between two species. We assume that the (first-order) rate coefficients (s$^{-1}$) for hopping and desorption are thermally activated according to the formula
\begin{equation}
R_X = \nu \exp\left(-\frac{E_X}{T}\right)
\label{RX}
\end{equation}
where $\nu$ is the pre-exponential factor, which is approximated by a constant value obtained from the transition state theory expression $kT/h \sim 2\times 10^{11}$ s$^{-1}$ at low temperature \citep[see Eq.~(4.45) in][]{Kolasinski:2002},  and $E_X$ is the barrier (or simple endoergicity) for  process $X$ in K. These energies can be constrained  by reproducing the laboratory results at different temperatures or obtained via \emph{ab-initio} calculations. Unlike desorption and hopping, the rate coefficients for those chemical reactions that have an activation energy barrier also have a tunnelling component.  In previous CTRW Monte Carlo studies \citep{Chang:2007,Cuppen:2007}, we treated reactions with barriers solely by tunnelling through a square potential and ignored any small temperature dependence at low temperatures. The present paper uses reaction rate coefficients that have a slight temperature dependence,  which were obtained in \citet{Fuchs:2009} by using the rate expression for thermal activation (Eq.~\ref{RX}) and allowing the activation energy barrier for reaction to be temperature dependent.  
Table~\ref{Energies} lists the resulting barrier heights and reaction rate coefficients for the reactions H + CO and H + H$_{2}$CO at temperatures from 12.0 K to 16.5 K.  The barriers  increase by $\sim$~20\% from the lowest to the highest temperature.   Since the intermediate products, HCO and H$_3$CO, were not experimentally observed, it was concluded that the hydrogenation reactions of these species have a negligible barrier. In the program, these two reactions proceed without any barrier at all. After a reaction, the formed product gains some energy to rearrange and it is allowed to make a few diffusion steps. 

\begin{table}[h]
\caption{Reaction rate coefficients, $R_{\rm react}$, and activation energy barriers for CO + H and H$_2$CO + H for different temperatures.
\label{Energies}}
\begin{center}
\begin{tabular}{ccccc}
\hline
$T$ & \multicolumn{2}{c}{CO + H} & \multicolumn{2}{c}{H$_2$CO + H}\\
& barrier &  $R_{\rm react}$ & barrier & $R_{\rm react}$ \\
(K) & (K) & (s$^{-1}$) &(K) & (s$^{-1}$)\\
\hline
12.0 & 390 $\pm$ 30 & $2 \times 10^{-3}$ & 415 $\pm$ 30 & $2 \times 10^{-4}$ \\
13.5 & 435 $\pm$ 40 & $2 \times 10^{-3}$ & 435 $\pm$ 40 & $2 \times 10^{-3}$\\
15.0 & 480 $\pm$ 50 & $3 \times 10^{-3}$ & 470 $\pm$ 50 & $5 \times 10^{-3}$ \\
16.5 & 520 $\pm$ 60 & $4 \times 10^{-3}$ & 490 $\pm$ 60 & $2 \times 10^{-2}$ \\
\hline
\end{tabular}
\end{center}
\end{table}

The desorption, or binding, energy for an adsorbate 
depends on an
energy parameter $E$.   To understand the role of this parameter, let us first consider a CO molecule either in the ice or adsorbing onto the surface.   The CO molecule is assumed to lie or stick in a configuration close to the $\alpha$-CO structure \citep{Vegard:1930}. In this configuration, each CO molecule has up to 14 neighbours to which hopping can occur, four in the same layer, four one layer above and below and one two layers above and below. The binding energy $E_{\rm bind}$ is determined by the additive energy contributions of the occupied neighbouring sites. The contributions are
$2E$ for the layers below and, if applicable, $E$ for the neighbours in the same layer (lateral bonds) or in upper layers.  {An alternative treatment for sites below the particle is to add 
a single overall contribution for longer range interactions from the ice layer. In this way, the added long range contribution is roughly the same for all species of the same kind, regardless of their very local environment. However, if a very porous structure forms, the lower neighbouring sites are most likely not all occupied and the added long range contribution is less, reflecting the `real' long range contribution. }  As an example, if a CO molecule lands on top of a CO layer in a multi-layered CO ice, its binding energy is $ 1(2E_{\rm CO-CO}) + 4(2E_{\rm CO-CO})  = 10E_{\rm CO-CO}$, while if it lies in a deeply embedded layer in the ice, its binding energy is the full $1(2E_{\rm CO-CO}) + 4(2E_{\rm CO-CO}) + 4(E_{\rm CO-CO}) + 4(E_{\rm CO-CO}) + 1(E_{\rm CO-CO}) = 19E_{\rm CO-CO}$.  From \emph{ab-initio} calculations (Andersson, in prep) the parameter $E_{\rm CO-CO}$ for CO is 63~K, so that the binding energy of a CO adsorbate onto CO ice is 630~K.  Experimentally a desorption energy of CO from a CO {surface} was determined to be $855 \pm  25$~K, which corresponds to 13-14$E_{\rm CO-CO}$.
For atomic hydrogen, $E_{\rm H-CO}$ was calculated  to be 32~K onto CO ice {(Andersson, in prep)}. 
For H$_2$ a value of $E_{\rm H_2-CO}=33$~K is used, {slightly higher than for H as expected from calculations of \cite{Hornekaer:2005} for H$_2$ and \cite{Al-Halabi:2006} for H atoms.} If lattice sites are occupied by heavier species than CO (mainly CH$_{3}$OH) the total binding energy of an H or CO, and the structure are assumed not to change.  If lattice sites are occupied by H atoms rather than CO, the lower value of $E_{\rm X-H}$ is used for those  relevant sites, so that the total binding energy for a specific CO is an expression containing both the parameters for CO and for H.   For atomic H, the van der Waals interaction with another atomic H in an adjoining site is rather small, \emph{i.e.},  $E_{\rm H-H} = E_{\rm H-CO}/10$. The energy parameters $E_{\rm X-CO}$ for H$_2$CO and CH$_3$OH and the intermediates, HCO and H$_3$CO, are chosen such that these species neither desorb nor diffuse (see below) from the grain at the temperatures studied. All adopted values for $E$ are summarised in Table \ref{E}.

\begin{table}
\caption{The energy parameter $E$ for the different species.}
\label{E}
\begin{tabular}{lrr}
\hline
Species & \multicolumn{1}{c}{H}     &  \multicolumn{1}{c}{CO} \\
 & \multicolumn{1}{c}{(K)}     &  \multicolumn{1}{c}{(K)} \\
\hline
H       &  3    &  32\\
H$_2$   &  3    &  33\\
CO      & 32    &  63\\
HCO     &160    &1600 \\
H$_2$CO &160    &1600\\
H$_3$CO &160    &1600\\
CH$_3$OH&160    &1600\\
\hline
\end{tabular}
\end{table}

The same parameter $E$ is used to help determine the hopping barrier from
site $i$ {to an empty adjacent site, $j$}.  If sites $i$ and $j$ have the same binding energy, then the barrier for hopping is  given by the equation
\begin{equation}
E_{\rm hop}(i,j) = \xi E, 
\label{Ehop1}
\end{equation}
where $\xi$ is a parameter that determines how high the barrier is compared with the binding energy.  
 Here $\xi = 8$ is used, so that the barrier height is  80\% of the binding energy of a surface adsorbate (see above).  As reported in \cite{Fuchs:2009}, the model results are relatively insensitive to the exact value of this parameter. In general, the effect of the diffusion rate was found to become larger with temperature, where faster diffusion results in less effective hydrogenation, since the residence time per site decreases and more H atoms will desorb. The value of $\xi = 8$ was chosen to result in acceptable simulation times and to reproduce the desorption energy/hopping barrier ratio  found by \cite{Katz:1999} for H and H$_2$ on silicates and amorphous carbon. The same parameter $\xi$ is used for all different species: since it is multiplied by the species-specific parameter $E$, the difference in hopping rates between species is automatically accounted for. 
 For this choice of $\xi$, H and H$_2$ are mobile for $T > 8$~K and CO for $T > 16.5$~K. Diffusion from low to high binding sites is possible from even lower temperatures. 
 
 If the binding sites $i$ and $j$ have different binding energies, we add a second term to the formula for  the hopping barrier in order to have the forward and backward rates obey detailed balance.  The overall expression then becomes
\begin{equation}
E_{\rm hop}(i,j) = \xi E + \frac{\Delta E_{\rm bind}(i,j)}{2},
\label{Ehop}
\end{equation}
where $\Delta E_{\rm bind}(i,j)$ is the difference in binding energy between the two sites.  
Diffusion into the structure of the ice is included as well. Minimum energy path
calculations suggest that CO and H can swap positions, enabling an adsorbing
H-atom to penetrate into the CO ice (Andersson, in prep). The barrier for
this process strongly depends on the layer in which the H-atom is
situated. In the simulations the barrier for this event is (350 +
5($z_1 + z_2$))~K for an H-atom to swap between layers $z_1$ and
$z_2$, where $z_i$ stands for the depth of the atom in ML. {This swapping mechanism was only found to occur for the top few layers, where there is more space to accommodate the movement involved in the swap (Andersson, in prep). The minimum energy path calculations do not include thermal effects, which could promote this mechanism by lowering the barrier in the upper layers. In the current implementation, the layer dependence of the swapping barriers effectively limits the swapping process to the top few layers. Deeper into the ice, the swapping mechanism is not thermally accessible. } Hydrogen diffusion throught the ice via interstitial sites was found not to reproduce laboratory results and is therefore not included.

Thus, each species can in principle undergo up to 15 processes: thermal desorption and hopping to one of the up to 14 neighbouring sites, or, if sites are occupied, reaction with the species in that site. Reactions without activation energy (\emph{e.g.}, H + HCO $\rightarrow$ H$_{2}$CO) occur with 100\% efficiency, whereas for reactions with barriers, we must simulate the competition between the efficiency of reaction and that of hopping out of the adjacent lattice sites. Some of these processes can be forbidden. Bulk species are, for instance, not allowed to desorb and not all species react with each other. The rate coefficients of all these processes are determined using Eq.~\ref{RX} and the barriers as discussed above. In this way, diffusion and reaction are automatically treated competitively.  

In addition to reactions occurring via the Langmuir-Hinshelwood mechanism, which happen as a result of diffusion into adjacent sites, the program includes reactions occurring via the Eley-Rideal mechanism, in which a gas-phase species lands atop a surface species and reacts with it.  The barriers for reactions are assumed to be independent of mechanism.  In the diffusive case, however, if two reactants do not tunnel under or cross over the activation energy barrier, they have multiple additional opportunities until they diffuse away from one another.  In the Eley-Rideal case, we allow only one opportunity.  For example,   an H atom landing directly atop a CO molecule gets one chance to overcome the reaction barrier and react. If the reaction does not proceed, the H atom can then hop further to find another reactant.  

Deposition onto a surface site occurs with a rate coefficient $R_{\rm dep}$ according to 
\begin{equation}
R_{\rm dep} = \frac{v_A n_{\rm A}}{4\rho},
\end{equation}
where \(n_{\rm A}\) is the absolute gas abundance of species A, \(\rho\) is the surface site density, and \(v_{\rm A}\) is the mean velocity of species A:
\begin{equation}
v_{\rm A} = \sqrt{\frac{8 kT_{\rm gas}}{\pi m_{\rm A}}},
\end{equation}
with \(T_{\rm gas}\) the temperature of the gas and $m_{\rm A}$ the mass of A.  We use the standard value of  \(\rho = 1 \times 10^{15}\) cm\(^{-2}\)
which agrees with the site density of the $\alpha$-CO (110) crystal face. 
During the simulation H, CO and H$_2$ are allowed to deposit on the surface. We use a molecular hydrogen abundance that is only slightly higher than the atomic hydrogen abundance to maintain a coverage of H$_2$ on the surface within a reasonable simulation time. 
The presence of H$_2$, which is also formed on the surface,  has an effect on the penetration of the atoms and the diffusion. 
The CO gas-phase abundance is adjusted to account for freeze-out during the simulation. This freeze-out rate (loss rate of gaseous CO), which is the deposition rate onto a surface site multiplied by the number of sites per grain and the grain concentration, depends upon the dust-to-gas number ratio for fixed gas-phase density $n_{\rm H}$.  
 This paper uses two values for this parameter: $1 \times 10^{-12} n_{\rm H}$ and $2 \times 10^{-12} n_{\rm H}$. The influence of the rate of depletion, \emph{i.e.}, the rate of disappearance of CO from the gas, is discussed in Section \ref{dust-to-gas}.  Gas-phase chemistry is not included here.

\section{Results}
\label{results}
\subsection{Ice abundances as functions of time}
In order to follow the build up of CO, H$_2$CO, and CH$_3$OH in the ice mantle, simulations at different temperatures and densities were carried out. All simulations commence with the same initial bare surfaces consisting of lattices with $50 \times 50$ sites and some simple surface steps (similar to Surface c in \cite{Cuppen:H2}) 
The choice of the initial surface (either silicates or amorphous carbon)) was found  to be unimportant after the build-up of a few monolayers of ice.  To start the simulation on a bare surface, the energy parameters used were chosen to be the same as for the CO surface.
All results were converted to grains with a standard size of 0.1~$\mu$m. The simulation surface is a square flat surface. The spherical, continuous nature of a grains is mimicked by periodic boundary conditions.
The size of the simulated surface has been found to be in the regime where no size effects are observed \citep{Chang:2005}; this regime is associated with surface abundances of reactive species such as H  larger than one per grain. For H$_2$ formation, a size dependence is observed, since for smaller grains the average surface density of atomic hydrogen is much smaller than one (accretion limit), but because of the stronger sticking of CO and H$_2$CO with respect to H atoms, the size dependence is probably only important for higher temperatures where CO and H$_2$CO start to desorb.

\begin{figure}[h]
\includegraphics[width=0.45\textwidth]{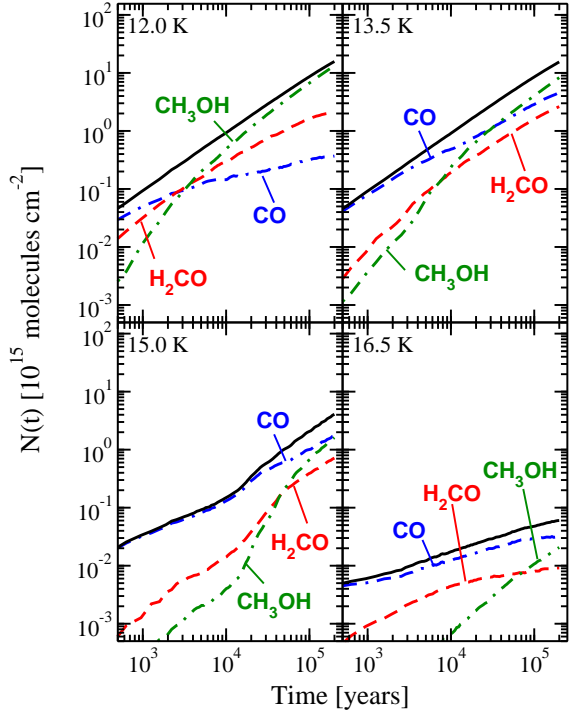}
\caption{CO, H$_2$CO, and CH$_3$OH build-up as a function of time for a density of $n_{\rm H} = 10^4$~cm$^{-3}$, $n_{\rm grain} = 2\times10^{-12} n_{\rm H}$  and $n({\rm CO})/n({\rm H}) = 1$ at four different surface temperatures (a) 12.0, (b) 13.5, (c) 15.0 and (d) 16.5 K. The black line indicates the total ice thickness, \emph{i.~e.~}the sum of the CO, H$_2$CO, and CH$_3$OH abundances.}
\label{1e4_1_1}
\end{figure}

\begin{figure}[h]
\includegraphics[width=0.45\textwidth]{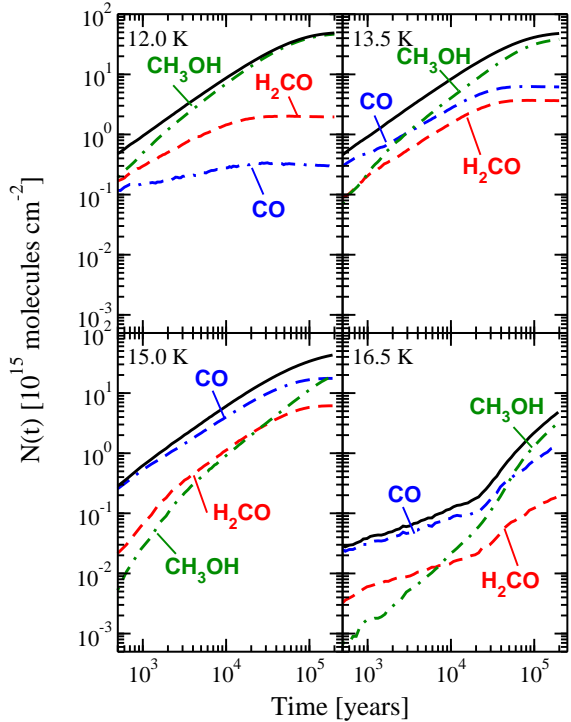}
\caption{Similar to Figure \ref{1e4_1_1} for $n_{\rm H} = 10^5$~cm$^{-3}$.}
\label{1e5_1_1}
\end{figure}

Figure \ref{1e4_1_1} shows the evolution of the three dominant ice species as a function of time for $n_{\rm H} =1 \times 10^{4}$~cm$^{-3}$, $n_{\rm grain}= 2 \times 10^{-12} n_{\rm H}$ and four different temperatures: 12.0, 13.5, 15.0, and 16.5 K. These temperatures are the same as used in the experiments in \cite{Fuchs:2009}. The abundances are plotted in terms of 10$^{15}$ molecules per cm$^{-2}$, which is equivalent to species per lattice site.   The plotted curves are the combined results of several independent simulations using different initial seeds, \emph{e.g.}, initial settings of the random number generator. 
Simulations were added until the evolution did not significantly change. 
Initial gas phase abundances of $n({\rm CO}) = 1 \times 10^{-4} n_{\rm H}$ and $n({\rm H}) = 1 \times 10^{-4} n_{\rm H}$ were assumed; these correspond to observed (CO) \citep{Lacy:1994} and calculated (H) fractional abundances for cold dense cores of a density of $n_{\rm H} =1 \times 10^{4}$~cm$^{-3}$. The abundance of atomic H is mainly determined by the ration between the cosmic-ray destruction rate of H$_2$ and the formation rate of H$_2$ which leads to $\sim 1$~cm$^{-3}$ independent of the density \citep{Duley:1984}. \cite{Goldsmith:2005} showed observationally that the H-atom abundance is higher than $1$~cm$^{-3}$ and we therefore use a higher H abundance of $10$~cm$^{-3}$ for our high density case which will be discussed below. The gas-phase abundance of CO is allowed to deplete as CO accretes onto grains while the gas-phase abundance of H remains constant. 
The solid line indicates the combined ice build-up of the three species together; in 10$^{5}$ yr, at most 10 monolayers of ice are produced at the value of $n_{\rm H}$ used.
From the graphs, a difference in this total build-up for the higher temperatures is immediately clear. Whereas the ice thickness grows linearly for 12.0 and 13.5~K, a clear non-linear behaviour as well as a drop in the total coverage can be observed at 15.0 and especially 16.5~K. This is due to the binding energy of CO. The residence time of an adsorbed CO molecule is 0.3~yr on a flat grain at 16.5 K. This time will increase once the CO molecules can stick together and form small islands, because the binding within the islands is stronger than for individual molecules. With the CO gas-phase concentration of 1~cm$^{-3}$, the arrival time of new CO molecules onto a grain  is of similar order as the residence time on a bare surface, so that the CO molecules do not have the opportunity to meet and stick together to form small islands, which is an efficient mechanism for growth.  Figure \ref{1e5_1_1} shows similar time evolution curves for a density of $n_{\rm H} =1 \times 10^{5}$~cm$^{-3}$, leading to a ten times higher flux, and indeed here there is a greater ice build-up  at 16.5 K. 

In general, the production of H$_2$CO and CH$_3$OH decreases with increasing temperature for both densities. This dependence is in agreement with the laboratory experiments of the hydrogenation of a CO ice {by \cite{Fuchs:2009}}. There the initial production rate was observed to be higher for low temperature whereas the final yield at the end of the experiments peaked at 15.0~K due to an increase of the penetration depth of H-atoms into the CO ice with temperature. In the co-deposition simulations presented in this paper, the penetration depth has less impact on the formation of H$_2$CO and CH$_3$OH since a fresh supply of CO is constantly deposited and the formation rate is therefore more comparable to the rate at the start of the experiments when a pure CO ice is exposed to atomic hydrogen.  In this condition, it is the shortened grain lifetime of H atoms at the higher temperatures that reduces the rate of methanol formation from CO. The rate for reaction changes only minimally with increasing temperature. 

The pure CO build-up here peaks at 13.5~K for both densities as can be clearly seen in Figs.~\ref{1e4_1_1} and \ref{1e5_1_1} by comparing the dash-dotted curves. This quantity is influenced by two effects with opposite temperature dependence:  the desorption of CO molecules (see above), which increases with increasing temperature, and the conversion of CO into H$_2$CO, which decreases with increasing temperature because of the lessened ability of H atoms to remain on grains long enough to react. Nevertheless, for most conditions, CH$_3$OH dominates the ice layer at late times and exhibits a very steep formation curve. CO and H$_2$CO on the other hand start with a high abundance at early times and increase much more slowly, even reaching a steady-state level on some occasions. The steady-state can be clearly seen for $T=12.0$ and 13.5~K for $n_{\rm H} = 1 \times 10^{5}$~cm$^{-3}$ and $T=16.5$~K for $n_{\rm H} = 1 \times 10^{4}$~cm$^{-3}$. 
For lower atomic H abundances, $\sim 1$~cm$^{-3}$, the H$_2$CO and CH$_3$OH formation is expected to occur at later times, since similar $n($H$)/n($CO) ratios will occur at later times. 

\subsection{Dust-to-gas number ratio}
\label{dust-to-gas}
The dust-to-gas number ratio determines the rate of the gas phase depletion and the maximum ice thickness that can be achieved before depletion of gaseous CO, which is roughly 50~ML for $n_{\rm grain}= 2 \times 10^{-12} n_{\rm H}$ and 100~ML for $n_{\rm grain}= 1 \times 10^{-12} n_{\rm H}$ as is discussed in the next section.

\begin{figure}[h]
\includegraphics[width=0.45\textwidth]{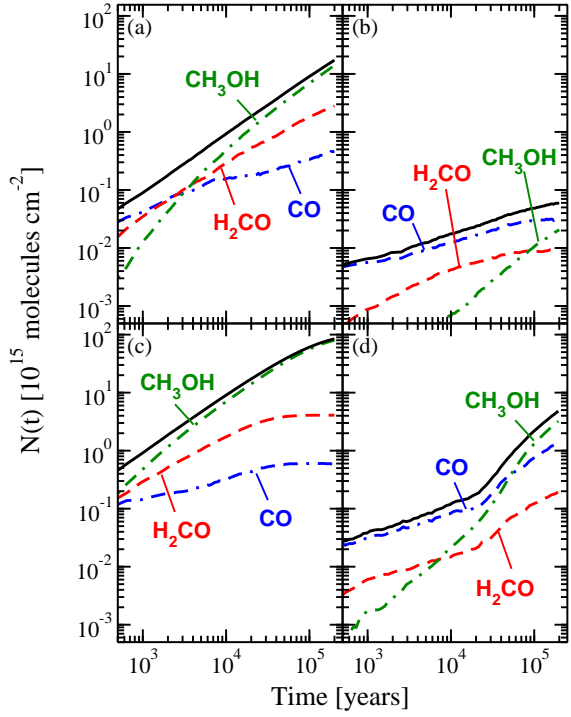}
\caption{CO, H$_2$CO, and CH$_3$OH build-up as a function of time for $n_{\rm grain} = 1\times10^{-12} n_{\rm H}$ and $n({\rm CO})/n({\rm H}) = 1$. (a) $T=12.0$~K and $n_{\rm H} = 10^4 $~cm$^{-3}$, (b) $T=16.5$~K and $n_{\rm H} = 10^4 $~cm$^{-3}$, (c) $T=12.0$~K and $n_{\rm H} = 10^5 $~cm$^{-3}$, and (d) $T=16.5$~K and $n_{\rm H} = 10^5 $~cm$^{-3}$. }
\label{1_1_2x}
\end{figure}

Figure \ref{1_1_2x} shows the temporal evolution of CO, H$_2$CO and CH$_3$OH  for 12.0 and 16.5~K with a reduced  dust-to-gas ratio of $1 \times 10^{-12} n_{\rm H}$. Densities of $n_{\rm H} = 10^4 $ and $10^5 $~cm$^{-3}$ are used. Panel (a)  can be compared with the 12.0~K panel in Fig.~\ref{1e4_1_1}a, panel (b) with the 16.5~K panel in Fig.~\ref{1e4_1_1} and  panels (c) and (d) with the 12.0 and 16.5~K panels in Fig.~\ref{1e5_1_1}, respectively. It is  apparent that only Fig.~\ref{1_1_2x}c is significantly different from its analogs.  
For the lower density cases ($n_{\rm H} = 10^4 $~cm$^{-3}$), substantial depletion of gaseous CO has not started yet after $2 \times 10^{5}$~yr, resulting in roughly the same gas-phase composition throughout the simulation for both dust-to-gas ratios. This is generally not true for the higher density models.  At 16.5~K, however,  the build-up of ice layers  is  hampered by the desorption of CO back into the gas phase, again resulting in very similar accretion rates for the two dust-to-gas cases.  At lower temperatures, however, the accretion rate for CO is greater for the lower dust-to-gas case. The main difference between Figs.~\ref{1e5_1_1}a (12.0~K) and \ref{1_1_2x}c is indeed that the levelling off to constant values for CO and H$_{2}$CO occurs at later times and thicker ice layers for $n_{\rm grain}= 1 \times 10^{-12} n_{\rm H}$. 

\subsection{Ice structure}
\begin{figure}[h!]
\includegraphics[width=0.45\textwidth]{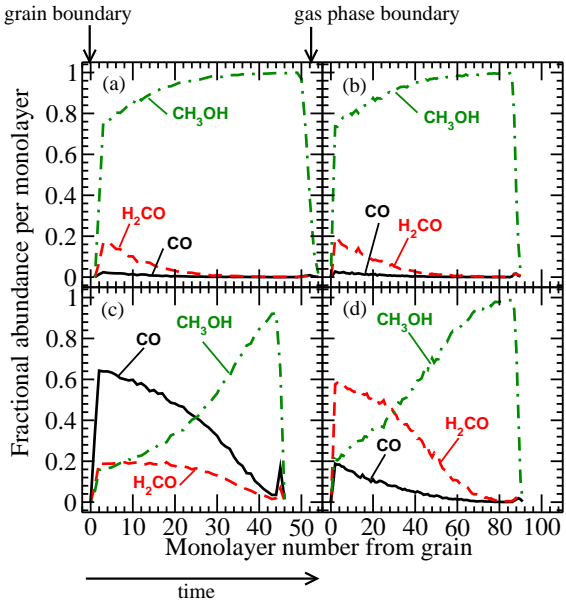}
\caption{Fractional abundance of the three main ice components as a function of monolayer after $2\times 10^{5}$~yr at $n_{\rm H} = 10^5 $~cm$^{-3}$ for (a)  12.0~K and $n_{\rm grain} = 2\times10^{-12} n_{\rm H}$, (b) 12.0~K and $n_{\rm grain} = 1\times10^{-12} n_{\rm H}$, (c) 15.0~K and $n_{\rm grain} = 2\times10^{-12} n_{\rm H}$, and (d) 12.0~K and $n_{\rm grain} = 1\times10^{-12} n_{\rm H}$ with a lower initial $n({\rm H})/n({\rm CO})$ gas phase abundance ratio of 0.5.}
\label{layers}
\end{figure}

In addition to the overall surface abundance as a function of time, the Monte Carlo approach allows us to obtain more detailed information concerning the layering of the ice.  Fig.~\ref{layers} plots the fraction per monolayer of the main ice components CO, H$_2$CO and CH$_3$OH at a time of $2 \times 10^{5}$ yr for 12.0 K (a, b, d)) and 15.0 K (c).  The abscissa is the number of the monolayer starting from the bare surface as zero {indicated by `grain boundary' in the figure. The label `gas phase boundary' indicates the top layer of the ice mantle, which faces the gas phase}.   Despite differences in dust-to-gas ratios and other parameters in the panels (see caption), the plot clearly shows that there is a gradient in the ice composition. At the onset of the CO freeze-out, a fraction of the layers remains in the form of CO and the H$_2$CO is not fully hydrogenated to CH$_3$OH. While the gas phase CO slowly depletes, the $n({\rm H})/n({\rm CO})$ ratio becomes more favourable for the complete hydrogenation of CO. For example, in the $2 \times 10^{5}$ yr of the simulation, the gas phase CO abundance drops from 10~cm$^{-3}$ to 0.2~cm$^{-3}$ for 12.0 K, $n_{\rm H} = 10^5 $~cm$^{-3}$ and $n_{\rm grain} = 2\times10^{-12} n_{\rm H}$. The right panels show the ice composition for slower CO depletion ($n_{\rm grain} = 1\times10^{-12} n_{\rm H}$). Panel (d) has additionally an altered  initial abundance of $n({\rm H})/n({\rm CO}) = 0.5$. Note that the horizontal scale  for the right panels differs from that for the left panels. For $n_{\rm grain} = 1\times10^{-12} n_{\rm H}$ the change to pure CH$_3$OH layers occurs over more layers than for $n_{\rm grain} = 2\times10^{-12} n_{\rm H}$. If panels (a) and (b) were plotted as a function of the fraction of the total ice thickness instead of the absolute ice thickness, very similar graphs would be obtained.  The final overall H$_2$CO/CH$_3$OH ratio after freeze-out is therefore independent of the exact dust-to-gas ratio. Comparison of panels (a) and (c) confirms that the rate and efficiency of the conversion of CO into methanol is determined by the temperature while comparison of panels (b) and (d) shows that the rate and efficiency are also determined by the $n({\rm H})/n({\rm CO})$ ratio.  A closer look at panel (d) compared with panel (b) shows that the lower
 $n({\rm H})/n({\rm CO})$ ratio leads to larger H$_2$CO and CO fractions for the lower monolayers  and that the conversion to pure methanol layers occurs only at higher layers, which are formed at later times.

\begin{figure}[h]
\includegraphics[width=0.45\textwidth]{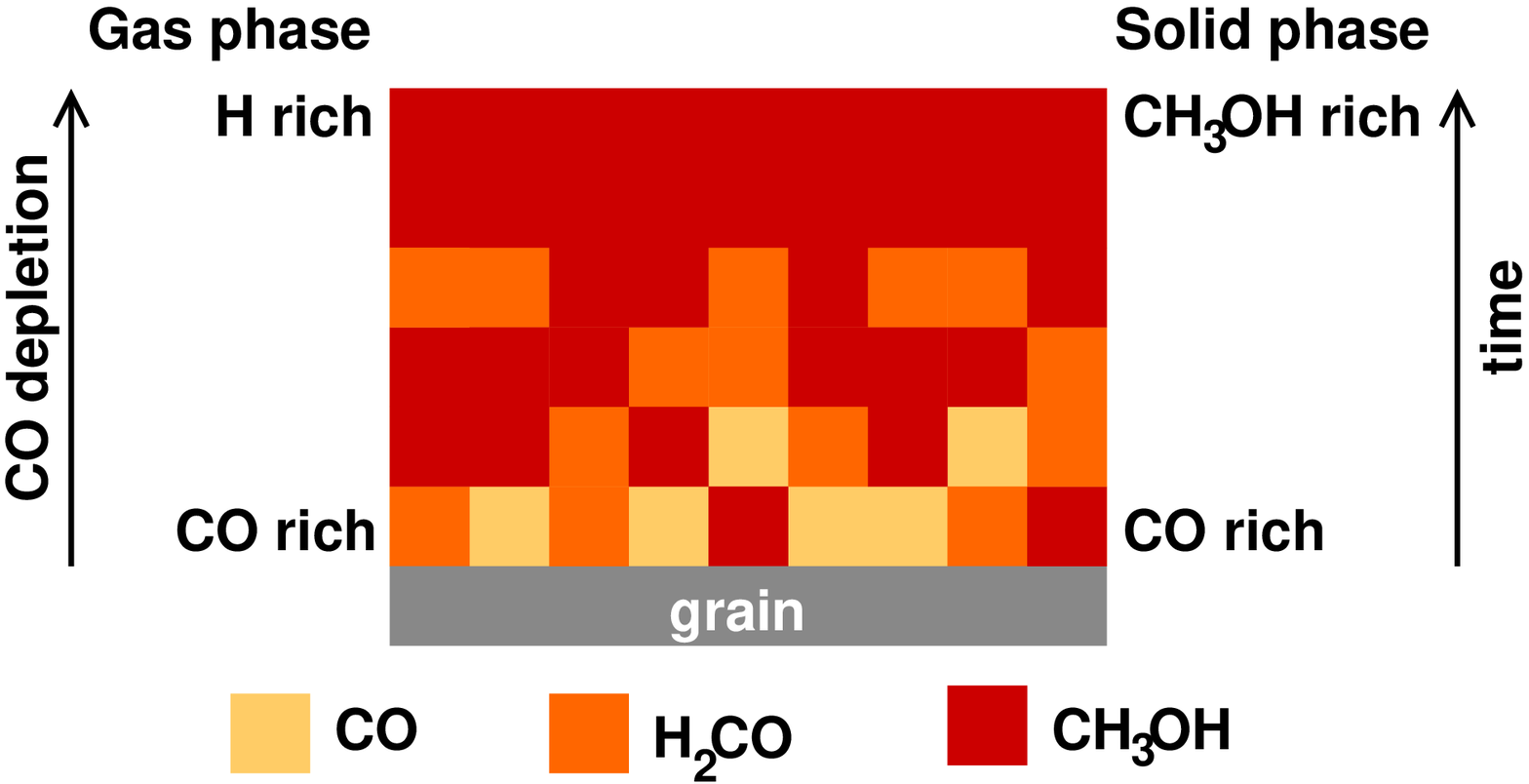}%
\caption{Schematic picture of the growth of the ice mantle during CO freeze-out. For coloured figures, see the online version.}
\label{schematic}
\end{figure}

Figure \ref{schematic} paints a schematic picture of a grain mantle. At the onset of CO freeze-out, the flux of CO molecules accreting onto the grain is large, larger than the part of H atom flux that is available for hydrogenation of CO, 
and, as a consequence, the mantle will be rich in CO with H$_2$CO and CH$_3$OH as minor components. As the $n({\rm H})/n({\rm CO})$ ratio in the gas phase increases, more and more CO and H$_2$CO is converted into CH$_3$OH and the outer layers of the mantle become more CH$_3$OH rich.
{Note that the astronomically observed solid abundances are sums over the entire grain mantle, although different ice components can be distinguished through the line profiles.}

\begin{figure}[h]
\begin{center}
\includegraphics[width=0.4\textwidth]{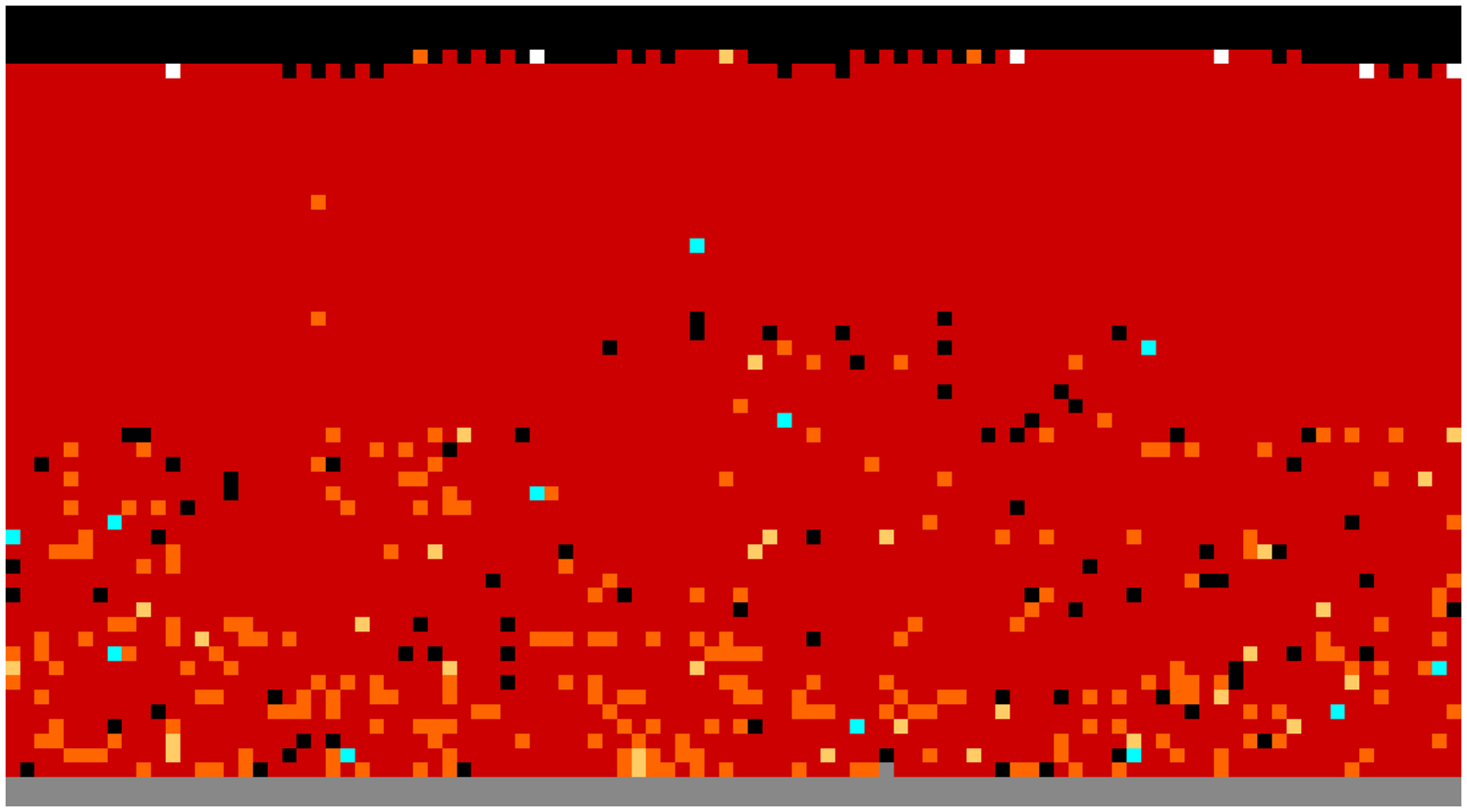}\\
\includegraphics[width=0.4\textwidth]{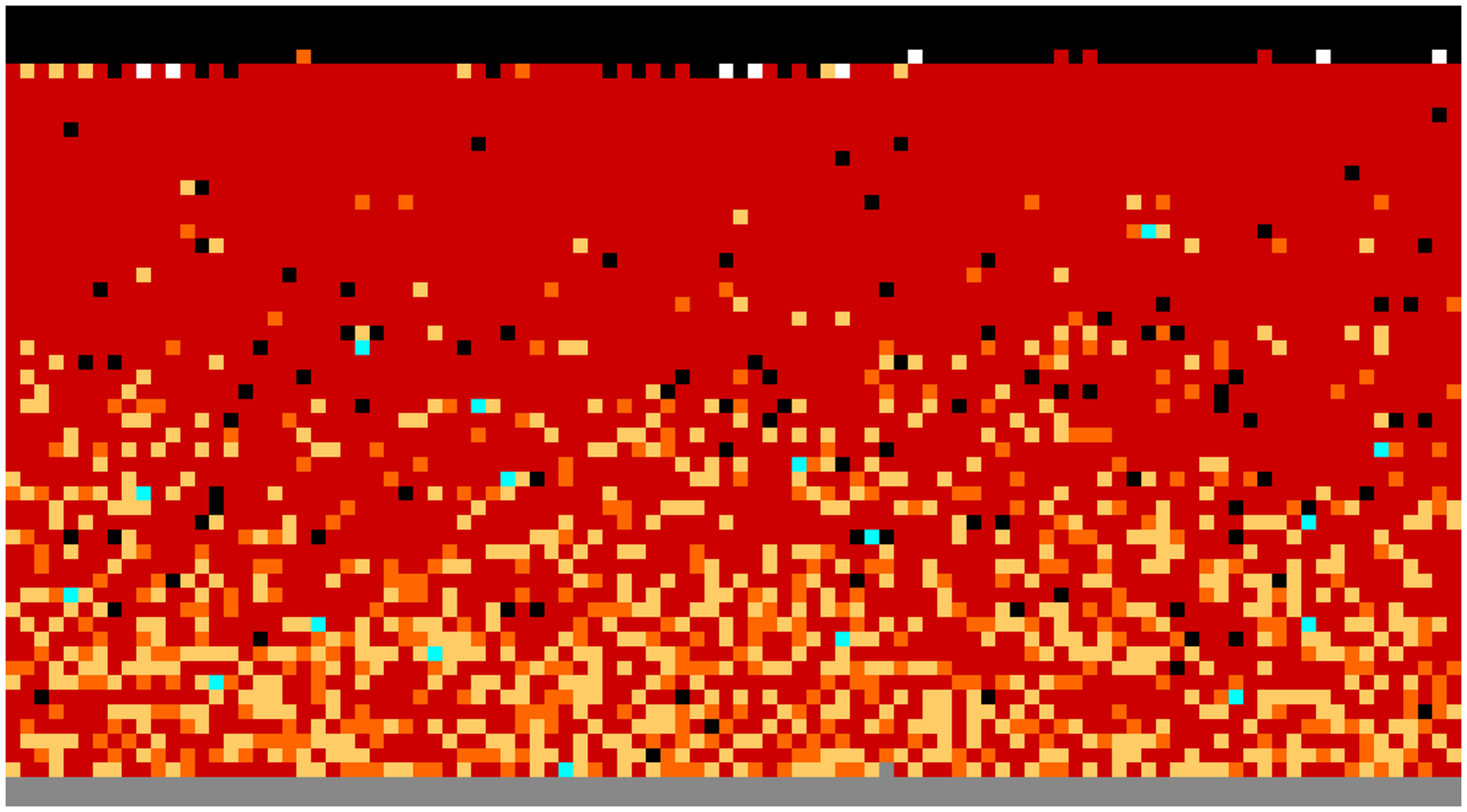}\\
\includegraphics[width=0.4\textwidth]{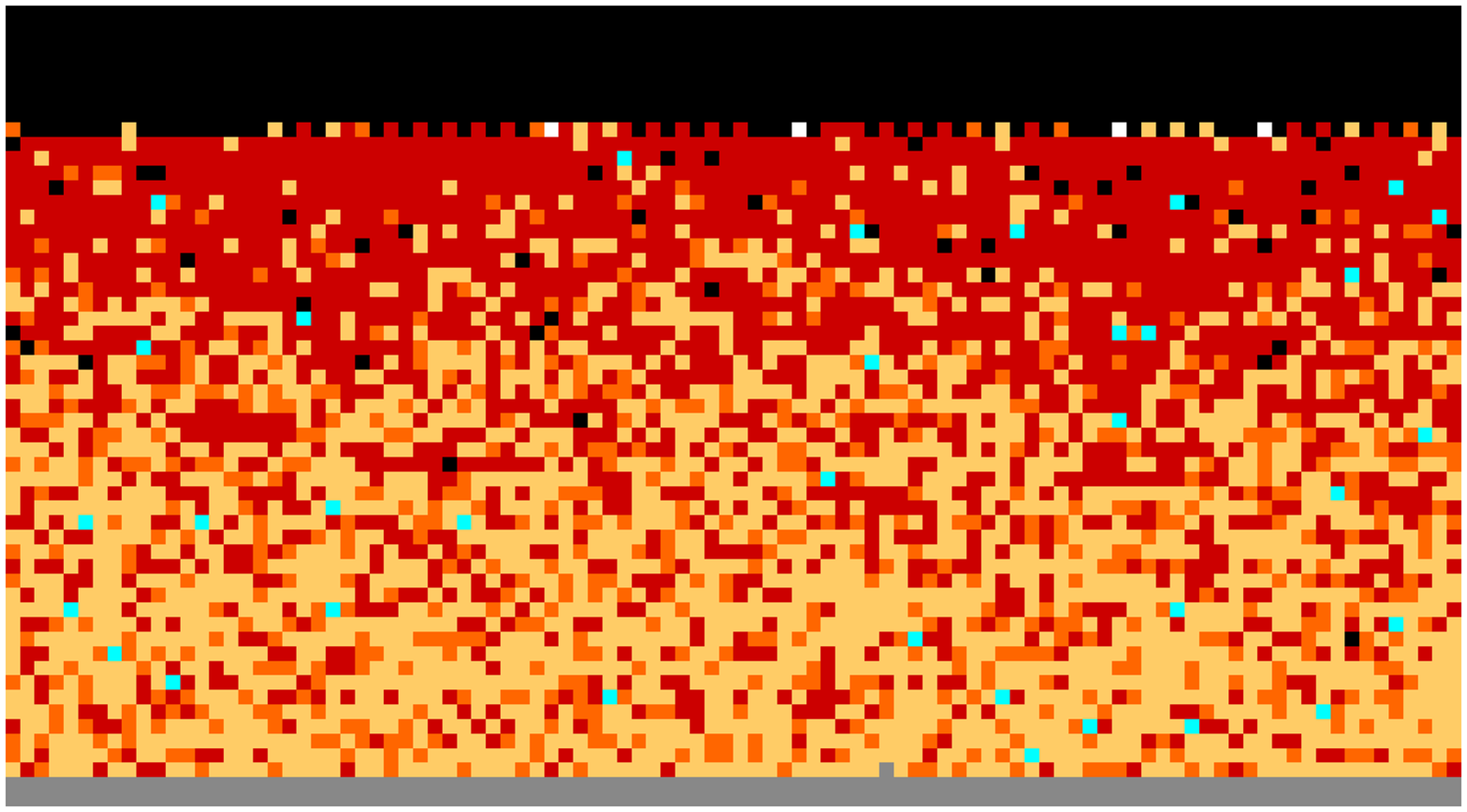}\\
\includegraphics[width=0.4\textwidth]{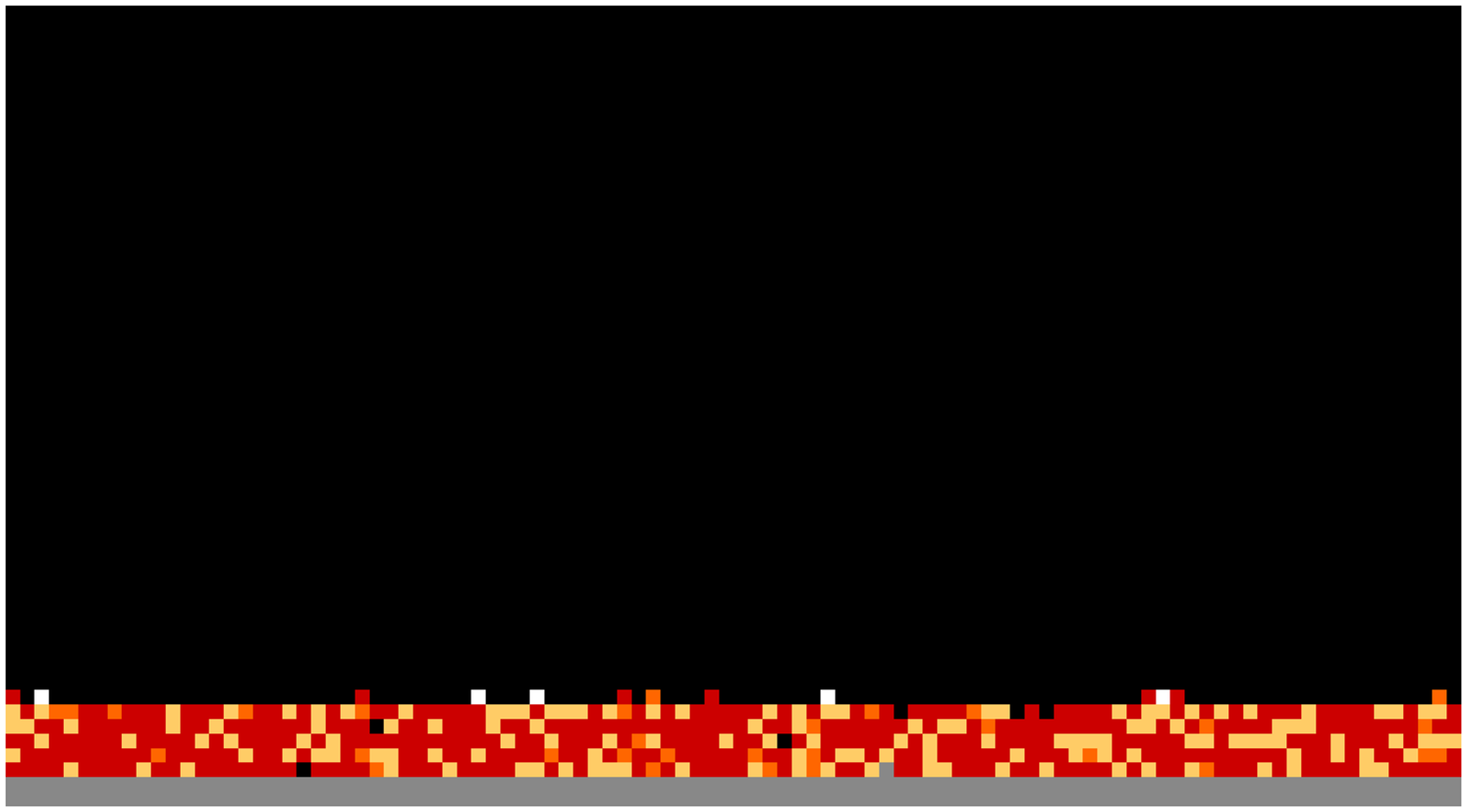}\\
\end{center}
\caption{Cross section of the ice mantle after $2\times 10^{5}$~yr at a density of $n_{\rm H} = 10^5 $~cm$^{-3}$ for 12.0, 13.5, 15.0 and 16.5~K (top to bottom). Unoccupied sites are indicated in black, grain by light grey, intermediates by cyan, and H and H$_2$ by white. CO, HCO, H$_2$CO, H$_3$CO, and CH$_3$OH are coloured by increasing darkness according to their degree of hydrogenation (see Fig~\ref{schematic}). For coloured figures,  see the online version.}
\label{surfaces}
\end{figure}

Actual simulated grain mantle cross sections for individual runs are plotted in Fig.~\ref{surfaces}. 
Here unoccupied sites are indicated in black, light gray on the bottom of the figures represents the bare grain, while H and H$_2$ are represented by white. All other mantle species, CO, HCO, H$_2$CO, H$_3$CO, and CH$_3$OH, have different Gray scale levels according to their degree of hydrogenation; \emph{i.e.},  CH$_3$OH is darkest, whereas CO is represented by the lightest Gray. The four panels correspond to the resulting icy mantle after $2\times 10^{5}$~yr under the same conditions as in Fig.~\ref{1e5_1_1}. The grain temperatures are 12.0, 13.5, 15.0 and 16.5~K  from top to bottom. The zigzagging pattern at the grain mantle surface reflects the zigzagging structure of $\alpha$-CO. The panels show clearly the gradient in CO, H$_2$CO and CH$_3$OH across the grain with the top layers containing the more saturated species. Furthermore, it shows that the lower temperature grains contain more CH$_3$OH. The grain mantle at 16.5 K is much thinner due to the desorption of CO at this temperature. Finally, some small pores can be observed in the mantles; these appear to form during hydrogenation where two species recombine to one, which takes up less space. It appears that at late times, when the flux is low, either CO or CH$_3$OH has some time to rearrange and fill most of these vacancies.

\subsection{Methanol formation at 8 and 10~K}
\label{cold}
\begin{figure}[h]
\includegraphics[width=0.45\textwidth]{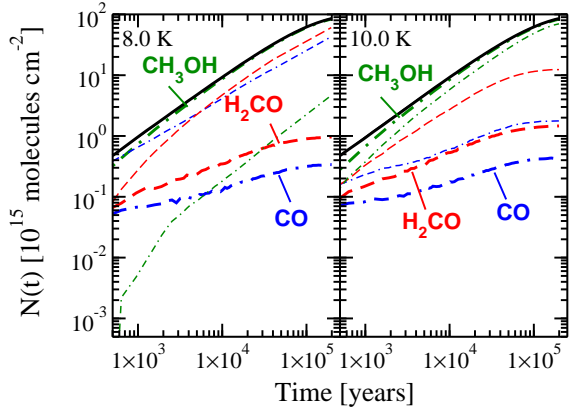}
\caption{CO, H$_2$CO, and CH$_3$OH build-up as a function of time for a density of $n_{\rm H} = 10^5$~cm$^{-3}$, $n_{\rm grain} = 1\times10^{-12} n_{\rm H}$  and $n({\rm CO})/n({\rm H}) = 1$ at two different surface temperatures (a) 8.0, and (b) 10.0~K. The black line indicates the total ice thickness, \emph{i.~e.~}the sum of the CO, H$_2$CO, and CH$_3$OH abundances. The reaction rates are assumed to be either governed by tunneling (thick lines) or purely thermal (thin lines). }
\label{8K_10K}
\end{figure}

The results presented above all concern the formation of methanol in the temperature regime between 12.0 and 16.5~K. These temperatures are more representative for the outer regions of protostellar {envelopes}. In cold dense cores, the temperature is most likely lower. Figure~\ref{8K_10K} presents simulation results for 8 and 10~K. The model is extended to temperatures outside the regime studied in the laboratory, by assuming that the reaction {rates ($R_{\rm react}$ in Table \ref{Energies})} do not change with temperature, since they are dominated by tunnelling, whereas all other processes are thermally activated. The results in Fig.~\ref{8K_10K} clearly show that formaldehyde and methanol are still efficiently formed at these temperatures. Simulations with much lower reaction rates for H+CO and H+H$_2$CO, thermally activated using barriers of 390 and 415~K, respectively, still result in the formation of both H$_2$CO and CH$_3$OH, although with much lower abundances as indicated by the thin lines in Fig.~\ref{8K_10K}.  The abundance of formaldehyde is much larger in this case, since it is not efficiently converted into methanol. However, the rates at higher temperature suggest tunneling to be important and the abundances are therefore probably better represented by the thick curves.

\section{Comparison with other models and  observations}
\label{comparison}
This section compares the results of our Monte Carlo simulations with different models and with observations of H$_2$CO and CH$_3$OH abundances. In particular, our results are compared with a similar model with similar input parameters using a standard rate equation technique \citep[see, \emph{e.g.},][]{Ruffle:2000}  and with different models of various degrees of chemical and astrophysical complexity that are reported in the literature.  
Finally, the simulations are compared with a steady-state model proposed by \cite{Charnley:1997} that we have adjusted to take the changing CO gas-phase abundance into account.

\subsection{Comparison with rate equations} 
\label{rate equations}
\begin{figure}[h]
\includegraphics[width=0.45\textwidth]{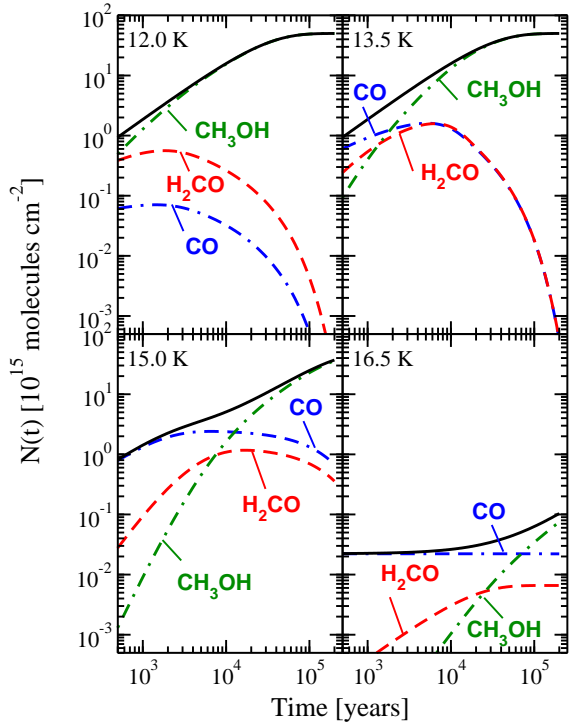}
\caption{Similar to Figure \ref{1e5_1_1} but using rate equations instead of Monte Carlo simulations.}
\label{rateeq}
\end{figure}

Figure \ref{rateeq} plots the results for a rate equation model \citep[see, \emph{e.g.},][]{Ruffle:2000} with the same chemical and physical processes as used in our Monte Carlo approach.  Besides the difference in mathematical techniques, the rate equation method uses single hopping and desorption rates rather than rates that depend on the local structure.  We expect the diffusion to be dominated by hopping between sites of the same type which results in a diffusion energy of $E_{\rm hop} = 8E$  (see Eq.~\ref{Ehop}) and the desorption dominated by strong binding sites (three horizontal neighbours), since diffusion will allow the particles to move to these sites where they remain attached.
 The competition between reactions with activation energy barriers and hopping and desorption is treated  by dividing the rate coefficient for reaction by the total rate including diffusion and desorption, as discussed in detail by \citet{Herbst:2008}.  The panels in Fig.~\ref{rateeq} can be compared with those in Fig.~\ref{1e5_1_1}, which are determined with the Monte Carlo approach. 

In comparison with the Monte Carlo results, two trends become immediately apparent in the rate equation results: (1) there is a clear drop in the CO and H$_2$CO abundances at late times and (2) the difference in the initial CH$_3$OH formation ($< 10^3$~yr) 
increases with surface temperature. The drop in the surface abundance of CO and H$_2$CO at late times is due to the mean field character of the rate equation technique. It treats all molecules of the same species in the same way, regardless of their position in the ice layers. CO ice that resides at the lower layers of the ice can in the standard implementation of the rate equation technique still react with impinging hydrogen atoms, whereas in reality this reaction will not proceed unless H atoms can penetrate into the porous ice.  Figure \ref{layers} shows that indeed most of the CO and H$_2$CO is buried deep into the ice. In the Monte Carlo simulations, the hydrogenation reactions will therefore be hampered by the deficiency of gas phase CO at late times whereas the rate equations continue hydrogenating the CO and H$_2$CO that have been formed at early times. In general, the effect of a changing gas phase composition is limited to the top layers of the ice. For this reason, layering should be taken into account when modelling the grain surface chemistry. The three-phase model introduced by \cite{Hasegawa:1993} is a first attempt to implement this effect into a rate equation model.

The second effect is probably due to the difference in the treatment of diffusion and desorption between the two methods. The agreement in the total ice thickness indeed becomes less if lower desorption barriers are used. Due to the implementation of the competition for reaction in the rate equation approach, the choice of the diffusion barrier has very little effect on the final results. This can be understood by realising that faster diffusion will lead to more encounters between diffusers and  the more stationary reacting species but that the residence time in the vicinity of the stationary reactant is reduced. The site dependent rates in the Monte Carlo code result in longer residence times at some sites and shorter residence times at other sites, favouring reactions, especially at higher temperatures, as first discovered for the formation of H$_{2}$ \citep{Cuppen:H2}.   { In addition to the two effects, it can also been seen that with the rate equation approach, the ice cannot even develop to 1 monolayer at 16.5~K, presumably because the CO growth mechanism of forming islands is not accounted for.}

\subsection{Comparison with other surface models}

\begin{figure}[h]
\includegraphics[width=0.45\textwidth]{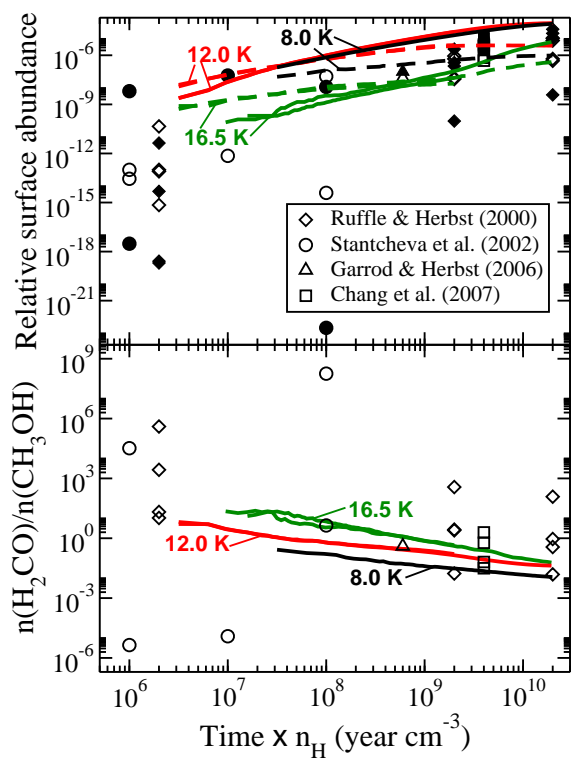}
\caption{Comparison of the present simulation data with values obtained by different models (see legend) as a function of time $\times$ density, which is a measure of fluence. The top panel displays the H$_2$CO and CH$_3$OH surface abundances relative to $n_{\rm H}$, the bottom panel H$_2$CO/CH$_3$OH ratios. Lines represent the present model data for 8.0 (black), 12.0 (red) and 16.5~K (green) and $n_{\rm H} = 10^4$ cm$^{-3}$ and $10^5$~cm$^{-3}$ (overlapping). H$_2$CO is in dashed lines and CH$_3$OH in solid lines. In the top panel, the open symbols represent H$_2$CO, the filled symbols CH$_3$OH. }
\label{models}
\end{figure}

A variety of  different models concerning the formation of formaldehyde and methanol via surface reactions from CO have been reported in the literature. Figure \ref{models} (top) shows fractional methanol and formaldehyde abundances with respect to $n_{\rm H}$ as functions of the product of time and density ($n_{\rm H}$) for a number of these models. This product is roughly proportional to the total fluence, the number of reaction species that reaches the grain surface, and can therefore serve as a measure to help compare models with different densities.
The solid lines indicate the methanol  and the dashed lines the formaldehyde abundances obtained in the present paper (Figs.~\ref{1e4_1_1}, \ref{1e5_1_1}, and \ref{8K_10K}). Only the data at 8.0, 12.0 and 16.5~K are plotted since the 12.0 and 16.5~K data represent the extreme values and 8.0~K is closed to cold core conditions. The $n_{\rm H} = 10^4$ and $10^5$~cm$^{-3}$ data overlap nicely indicating that the formation rate of methanol and that of formaldehyde are directly dependent on the fluence and that the flux difference of one order of magnitude does not introduce additional scaling effects. 

The symbols represent the fractional abundances of both species obtained from a selection of different studies. Open symbols represent H$_{2}$CO whereas filled symbols represent CH$_{3}$OH. Given the large differences in parameters, and assorted methods of calculation among the various studies, large variance in results can be expected.  The diamonds show results by \cite{Ruffle:2000}, who used  a large reaction network of both gas phase reactions and grain surface reactions. Both chemistries were treated using rate equations for four different scenarios defined by slow and fast diffusion rates combined with atomic and molecular initial conditions. 
The circles represent master equation results from \cite{Stantcheva:2002}. Here for a wide range of densities ($10^3$, $10^4$, $10^5$~cm$^{-3}$) the gas-grain chemistry for a limited set of surface reactions is obtained. Five additional reactions with respect to the surface network of the present paper were included, leading to the formation of H$_2$O, CO$_2$ and O$_2$. No gas phase chemistry was considered. 
The two triangles are obtained from \cite{Garrod:2006}, who followed both the gas and grain chemistry during the collapse and heat-up phase of hot cores using a rate equation approach. The points used here are in the early times of the collapse when the density is still close to the initial density. Again a full network was used, comparable to \cite{Ruffle:2000}, but with intermediate diffusion rates. In all three models \citep{Ruffle:2000,Stantcheva:2002,Garrod:2006} the barrier crossing of surface reactions with activation energy was treated simply by multiplying the meeting rates of reactive species  by the probability of crossing the activation energy  barrier. Unlike our rate equation approach (see above),  surface reactions with barriers were not treated to be in competition with other processes like diffusion \citep{Herbst:2008}. This neglect magnifies the differences obtained between low and high diffusion rates.  Furthermore, these three models do not capture the layered structure of the ice mantle but allow all species to react with each other, regardless of their position. This assumption can become a problem once multiple layers start to build up, as shown in the previous section.

The squares in Fig.~\ref{models} represent data from \cite{Chang:2007} obtained by a similar Monte Carlo method to that used in the present paper \citep{Chang:2005}, which includes both the layering effect as well as the competition of the reaction barriers. The surface chemistry is coupled to the gas phase chemistry and the surface chemistry network is very similar to one used by \cite{Stantcheva:2002}. The square symbols represent results for $n_{\rm H} = 2 \times 10^4$~cm$^{-3}$ and a surface temperature of 10 and 15 K. 
The main difference between the method by \cite{Chang:2007} and the present simulations is that our Monte Carlo algorithm is optimised to reproduce laboratory experiments and it includes important features of the CO + H system that were revealed through this optimisation. 

In general, Fig.~\ref{models} shows a large spread which reflects the different levels of complexity among the different methods. Most of the obtained abundances of  H$_2$CO and CH$_3$OH lie between our 12.0 and 16.5~K results and follow roughly the same trend that H$_2$CO dominates at early times and CH$_3$OH at late times. It appears that the slow diffusion results underestimate the formation of both molecules and the CH$_3$OH/H$_2$CO ratio, as compared to our model that is parameterized by  experiments. Our 8.0~K data lie on the high end of the results, probably for two reasons. Our $n({\rm CO})$ is relatively high throughout the entire simulation, although it should be comparable to the intermediate abundance of \cite{Stantcheva:2002}, and our limited network might overproduce H$_2$CO and CH$_3$OH instead of forming H$_2$O and CO$_2$ for instance. Our simulations are however geared towards conditions where most of the atomic oxygen is locked up in CO, and H and CO are indeed the most important reactants on the surface.

The bottom panel of Fig.~\ref{models} shows the H$_2$CO/CH$_3$OH ratio for the low temperature models ($< 20$~K) presented in the top panel. 
It is immediately evident from this plot that there is a spread in this quantity for the different models, even within the same paper.  The difference in the H$_2$CO/CH$_3$OH ratio for our 8.0, 12.0 and 16.5~K is much smaller compared with the spread found in other models. This spread is due to different assumptions in the modelling method and differences in methodology. In a few instances, the spread is due to different physical conditions such as temperature (as in our data) and density. In general, the largest differences with our results come from models with a low diffusion rate that do not consider the competition between reaction and diffusion. 

\subsection{Comparison with analytical model}
\cite{Charnley:1997} derived an analytical steady-state rate equation approach to predict the CO/CH$_3$OH and H$_2$CO/CH$_3$OH surface abundance ratios as functions of $\alpha_{\rm H}$, the relative H-to-CO flux ratio,  and  $\phi_{\rm H} = P_{\rm CO}/P_{\rm H_2CO}$ , the ratio of the probabilities for single CO and H$_{2}$CO  molecules to react with an H atom; \emph{i.e.}, the ratio of the rate coefficients. The exact derivation of these expressions is given in Appendix A.  Each value for $\alpha_{\rm H}$ and $\phi_{\rm H}$ results in a unique combination of both abundance ratios.  When \citet{Charnley:1997} was published,  $\phi_{\rm H}$ was unknown. Now, we can use Table~\ref{Energies} to obtain this quantity for different temperatures, so as to compare this steady state rate equation approach with the more realistic Monte Carlo simulations.

The top panel of Fig. \ref{Keane} plots the $n$(CO)/$n$(CH$_3$OH) abundance ratio versus the $n$(H$_2$CO)/$n$(CH$_3$OH) abundance for the Monte Carlo simulations (open symbols except for 16.5 K) and for the steady-state model (lines).  For each temperature,  $\phi_{\rm H}$ is indicated as obtained from Table~\ref{Energies}.  
The Monte Carlo curves follow the abundance ratios in time; \emph{i.e.}, they start from a relatively high CO gas-phase abundance which is partially converted into formaldehyde and methanol, so that the general direction is right to left.  Likewise, the general direction is from top to bottom since formaldehyde is generally converted into methanol.
The steady-state model by \cite{Charnley:1997} plots curves for varying $\alpha_{\rm H}$. Since $\alpha_{\rm H}$ increases gradually during the course of our simulations, both approaches can be compared in a relative straightforward way  although the simulations have also other sources of time dependence.
The time it takes for the top layer of the grain mantle to reach steady state is short as compared to the change in $\alpha_{\rm H}$.
Figure \ref{Keane}(top) includes the steady-state model lines for values of $\phi_{\rm H}$ that correspond to the values in the simulations. We would like to emphasise that the lines by \cite{Charnley:1997} represent $\theta_{\rm CO}$/$\theta_{\rm CH_3OH}$ and $\theta_{\rm H_2CO}$/$\theta_{\rm CH_3OH}$, where $\theta_{\rm X}$ indicates the steady-state coverage of species X in the top layer, whereas the abundances in the Monte Carlo simulations are for the entire grain mantle. At first sight, very good agreement is obtained between the simulation results and the analytical model, especially considering the requirement for steady state in the latter, which is not fulfilled in the Monte Carlo simulations. However, it must be remembered that in the absence of a third dimension to the plot, $\alpha_{\rm H}$ is a hidden parameter, and the agreement for $\alpha_{\rm H}$ between the two methods for the same points on the figure is not good. Indeed, the same results on Figure \ref{Keane} are obtained for a 5-20 times lower H/CO flux in the steady-state model as compared with the simulations. In the simulations, a significant portion of the H atoms are `lost' to H$_2$ formation and desorption. Both processes are not included in the model by \cite{Charnley:1997}. 

The smaller symbols in the right top corner are the result of simulations with a different starting H abundance, chosen such that the flux of CO molecules is 10 times higher than the H atom flux, \emph{e.g.}, $\alpha_H $is $0.1$ at the start of the simulation. 
The absolute abundances of H and CO have little influence on the obtained ratios.
These independent simulations continue nicely on the analytical curve. As \cite{Keane:thesis} shows, the introduction of more competing hydrogenation reactions leads to different curves. Section \ref{comp reactions} addresses this point in more detail. 

One purpose behind a plot like Fig.~\ref{Keane} is to visualise whether H atoms are more likely to react with CO or H$_2$CO. The laboratory results indicate that for low temperature (12.0~K) CO + H is dominant whereas H$_2$CO hydrogenation becomes the preferred channel for higher temperatures (16.5~K). The plot shows that, even though the $\alpha_{\rm H}$ agreement is not very good, the agreement for $\phi_{\rm H}$ is very nice. 
The curves in Fig.~\ref{Keane}(top) are for one value of $\phi_{\rm H}$ and varying $\alpha_{\rm H}$, which can be compared with  the simulations at a single temperature. The steady-state model and Monte Carlo simulation results for the same value of $\phi_{\rm H}$ trace the same unique lines in the $n$(CO)/$n$(CH$_3$OH)  versus  $n$(H$_2$CO)/$n$(CH$_3$OH) plot. 
Since $\phi_{\rm H}$ decreases with temperature, plotting observational H$_2$CO/CH$_3$OH versus CO/CH$_3$OH ratios could give a good temperature and evolutionary indication between sources.

The main conclusions that can be drawn from these comparisons are that the surface chemistry of the top layer can at each point in time be very well approximated by quasi-steady state conditions, since the time to reach a quasi-steady state is smaller than the change in flux,
and that the majority of the hydrogen atoms do not react but leave the surface by desorption. This means that the conversion rate of CO and H$_2$CO into more saturated species is overestimated by the model of \cite{Charnley:1997}. 
Table~\ref{rates} lists the steady-state conversion fractions of CO into H$_2$CO and CH$_3$OH as a function of temperature and H/CO gas phase abundance ratio obtained from the Monte Carlo simulations.
The calculated fractions include the effect of desorption of atomic hydrogen and the competition of the CO hydrogenation reactions with the formation of H$_2$, which is not included in the analytical model.  

{These conversion fractions clearly show that CO is very efficiently converted into H$_2$CO and CH$_3$OH. Naturally, the exact numbers depend on many assumptions regarding for instance the values of $E$, $\xi$ and the reaction barriers. The parameter $E$ is central in determining the desorption temperature of the species and an increase in $E$ will therefore result in an appreciable H-atom abundance at higher temperature increasing the hydrogenation regime with a few degrees. The diffusion parameter $\xi$, on the other hand, will have a stronger effect at low temperature. Finally, a change in the reaction barriers will lead to a slightly different H$_2$CO/CH$_3$OH ratio.}

\subsection{Comparison with observations }
\begin{figure}[h]
\includegraphics[width=0.45\textwidth]{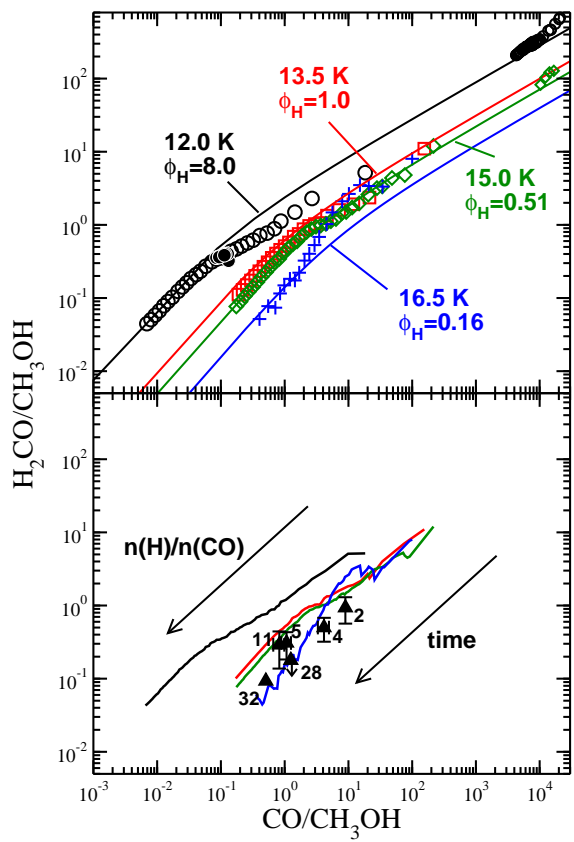}
\caption{CO/CH$_3$OH versus H$_2$CO/CH$_3$OH ice abundance. The top panel compares results as obtained by the Monte Carlo simulations (open symbols) and Eqs.~\ref{h2co/ch3oh}-\ref{co/ch3oh} (solid lines). The filled circles represent a simulation at 12.0 K that includes an O-atom flux. The bottom panel indicates the Monte Carlo results by solid lines. The triangles indicate ratios from ice observations where the numbers refer to the methanol content in the ice with respect to H$_2$O (see text). The quantity $\phi_{\rm H}$ represents the ratio between the H+CO and H+H$_2$CO reaction rates.}
\label{Keane}
\end{figure}

A comparison between the simulation results presented in this paper 
and observations is most straightforward with IR ice data. As long as the ice is not processed too severely, either by UV photons, which can break down formaldehyde and methanol, or by heating, which can sublimate CO, the IR data represents the most direct comparison with our simulations (see Section \ref{comp reactions} below). Ideal conditions are most likely found in cold dense cores or the outer cold envelopes of YSO's.   Since it is difficult to use absolute column densities for the comparison, the same ratios are used as in Fig.~\ref{Keane}. CO is usually present on grains in different mixtures: pure CO, polar CO and CO mixed with CO$_2$ \citep{Pontoppidan:2008}.  Laboratory spectra of CO mixed with either H$_2$O or CH$_3$OH show similar changes in the 4.7~$\mu$m CO feature, broadening and redshifting \citep{Bouwman:2007, Bisschop:thesis, Bottinelli:prep}. The polar CO component at 2139~cm$^{-1}$, normally ascribed to CO in a water-rich ice, can therefore also be due to a methanol-rich mixture. A CO:CH$_3$OH=1:1 mixture is already sufficient to shift the band to the observed range. Thus, to obtain the observed CO/CH$_3$OH ratio for comparison with our models, we use both the pure CO component and the polar CO component to account for the maximum amount of CO mixed with the formed CH$_3$OH ice. The triangles  in the bottom panel of Fig.~\ref{Keane} represent ratios obtained from ISO data for the YSO's AFGL 989, AFGL 2136, W33A, AFGL 7009S, and NGC 7538 IRS9 \citep{Gibb:2004} and VLT-ISAAC data for the Class 0 source Serpens SMM 4 \citep{Pontoppidan:2004}. Arrows indicate upper limits. As with the earlier data, Spitzer spectra of sources near SMM4 by \cite{Boogert:2008} resulted only in upper limits for H$_2$CO. They found formaldehyde to contribute perhaps 10-35\% of their C1 component, which cannot be assigned to a single species. \cite{Gibb:2004} used a feature equivalent to the C1 component to obtain their reported H$_2$CO abundances. Both the H$_2$CO abundances by \cite{Gibb:2004} and the H$_2$CO part of the C1 components by \cite{Boogert:2008} vary minimally with respect to water ice and are typically 6~\%. The range in H$_2$CO/CH$_3$OH abundances is therefore mainly due to methanol.
The abundance ratios for the six sources nicely overlap with the simulation data (solid lines). The good agreement with the CO/CH$_3$OH ratio could however be deceiving if part of the CO ice has already desorbed. 

Solid methanol abundances are found to vary substantially with respect to water ice, ranging from upper limits of a few percent to more than 30\% \citep{Dartois:1999,Pontoppidan:2003, Boogert:2008}. In Fig.~\ref{Keane}, these percentages are indicated  for the six sources compared here, spanning a range of more than an order of magnitude, assuming that, as discussed above, H$_{2}$CO ice does not vary much in abundance. A reason for this large range could lie in the differing evolutionary stages of the sources.
If most of the water ice is formed first before the catastrophic freeze-out of CO, from which methanol is formed, the methanol over water ratio will increase in time. The temporal evolution of the simulated abundance ratios proceeds from the top-right corner to the bottom-left corner of the figure, and the increase in methanol abundances for the six sources appears to follow this trend,
suggesting that indeed the large spread in methanol observations can simply be due to a difference in evolutionary stage of the outer envelope.  
Unfortunately, solid H$_2$CO has only been detected in a handful of sources and the same holds for clear upper limits for this molecule.

An alternative observational test could be to perform CO, H$_2$CO, and CH$_3$OH gas phase observations in very cold  and dense cores, where all three species are frozen-out onto the grains. Since most molecules have similar non-thermal desorption rates \citep{Oberg:photoI,Oberg:photoII}, one would expect the trace amounts in the gas phase to be representative of the ice layer composition \citep{Oberg:2009}.  The problem with this
technique is that not all observed gas-phase CO may result from CO ice evaporation and that gas-phase reactions can contribute to H$_2$CO as well. As shown in Figure 12 of \cite{Fuchs:2009}, the current model can reproduce the observed H$_2$CO/CH$_3$OH ratios in high-mass hot cores where the majority of the observed gas-phase molecules are likely evaporated ice species.

By ignoring the formation of water ice, our model predicts that CH$_3$OH is mainly present on the grain mantles in the pure form or mixed with CO, and that it is not in a water-rich phase. Unfortunately, the ice composition has very little effect on the 9.75~$\mu$m band profile of CH$_3$OH, which is usually used to determine its abundance, neither in peak shape nor position. The feature only becomes red-shifted when water ice is dominant, $>$90\% \citep{Skinner:1992, Bottinelli:prep}. Similarly, the 3.54 $\mu$m CH$_3$OH feature can be used to constrain the CH$_3$OH ice environment \citep{Dartois:1999, Pontoppidan:2003, Thi:2006}. All of these studies generally conclude that at least a fraction of the CH$_3$OH ice is in a water-poor, CH$_3$OH-rich environment but additional laboratory data on CH$_3$OH mixtures with CO are needed to quantify this.

\begin{table}
\caption{Conversion fraction of CO into H$_2$CO and CH$_3$OH
as a function of $T$ and H/CO
\label{rates}}
\begin{center}
\begin{tabular}{rl@{$\times$}ll@{$\times$}ll@{$\times$}ll@{$\times$}l}
\hline
$\frac{n({\rm H})}{n({\rm CO})}$  &
           \multicolumn{8}{c}{Temperature (K)}                  \\
     &        
           \multicolumn{2}{c}{12.0} &
           \multicolumn{2}{c}{13.5} &
           \multicolumn{2}{c}{15.0} &               
           \multicolumn{2}{c}{16.5}\\
\hline           
H$_2$CO \\
0.50 & $5.4$&$10^{-1}$ & $2.4$&$10^{-1}$ & $1.4$&$10^{-1}$ & $3.0$&$10^{-2}$ \\
0.75 & $3.7$&$10^{-1}$ & $2.3$&$10^{-2}$ & $1.7$&$10^{-1}$ & $3.0$&$10^{-2}$\\
1.00 & $1.8$&$10^{-1}$ & $2.1$&$10^{-1}$ & $1.9$&$10^{-1}$ & $4.0$&$10^{-2}$\\
2.00 & $6.8$&$10^{-3}$ & $4.3$&$10^{-2}$ & $1.6$&$10^{-1}$ \\
5.00 & $1.0  $&$10^{-5}$ & $1  $&$10^{-3}$ & $2.1$&$10^{-2}$ \\
CH$_3$OH\\
0.50 & $2.1$&$10^{-1}$ & $1.8$&$10^{-1}$ & $5.2$&$10^{-2}$ & $8.5$&$10^{-1}$ \\
0.75 & $5.4$&$10^{-1}$ & $3.1$&$10^{-1}$ & $1.1$&$10^{-1}$ & $7.5$&$10^{-1}$\\
1.00 & $9.9$&$10^{-1}$ & $3.3$&$10^{-1}$ & $1.1$&$10^{-1}$ & $7.1$&$10^{-1}$\\
2.00 & $1.0$&$10^{0}$  & $8.6$&$10^{-1}$ & $4.4$&$10^{-1}$ \\
5.00 & $1.0$&$10^{0}$  & $1.0$&$10^{0}$  & $1.0$&$10^{0}$\\
\hline
\end{tabular}
\end{center}
\end{table}

\section{Impact of competing reactions}
\label{comp reactions}
The Monte Carlo simulations reported in the previous sections all involve hydrogenation of pure CO ice. The reason for this is two fold. 
First, one aim of this study is to model CO freeze-out in the centre of cold cloud cores \citep{Pontoppidan:2006,Pontoppidan:2008} where most of the gas is in molecular form and H$_2$ and CO are the dominant species. In these centres,  most of the elemental oxygen is in the form of CO or frozen out into grains in form of H$_2$O. 

\begin{table}
\caption{Surface reactions in the H, O and CO system }
\label{reactions}
\begin{center}
\begin{tabular}{lcll}
\hline 
Reactants & &Products & $E_{\rm a}$(K) \\ \hline
H + H &$\rightarrow$& H$_2$ & \\
H + O &$\rightarrow$& OH &  \\
H + OH &$\rightarrow$& H$_2$O & \\
H + CO &$\rightarrow$& HCO & Table \ref{Energies} \\
H + HCO &$\rightarrow$& H$_2$CO & \\
H + H$_2$CO &$\rightarrow$& H$_3$CO & Table \ref{Energies} \\
H + H$_3$CO &$\rightarrow$& CH$_3$OH & \\
O + O &$\rightarrow$& O$_2$ &  \\
O + OH &$\rightarrow$& O$_2$ + H &  \\
 CO + OH &$\rightarrow$& CO$_2$ + H & 176 \\
 OH + H$_2$CO &$\rightarrow$& HCO + H$_2$O & \\
 OH + H$_3$CO &$\rightarrow$& H$_2$CO + H$_2$O & \\ \hline
\end{tabular}
\end{center}
\end{table}

To study the effect of competing reactions, however, simulations that include oxygen atoms have been performed as well. These simulations use a reaction network that is similar to that  reported by \cite{Chang:2007}, leading to the formation of O$_2$, H$_2$O, and CO$_2$. In addition, H$_2$CO and H$_3$CO can be destroyed by reaction with OH, leading to the formation of water (see Table \ref{reactions}).  All new species are assumed to bind very strongly and have only minimal diffusion. Eley-Rideal reactions are again allowed. The gas-phase abundances of CO and O are both assumed to be $5 \times 10^{-5} n({\rm H})$. Figure \ref{Keane} contains  results for the simulation at 12.0 K using filled circles. From this figure, two things are immediately clear: the time dependence of the CO/CH$_3$OH and H$_2$CO/CH$_3$OH ratios is minimal in the simulation  and the filled circles overlap nicely with the analogous simulations that do not include the extra reactions. \cite{Keane:thesis} presents a similar graph obtained with a full gas-grain network using macroscopic Monte Carlo simulations.  Their curves obtained for the full network are less steep than for the analytical expressions that are based on a limited network, since a considerable fraction of the hydrogen atoms reacts with other grain species. This is in contrast with our simulations that include H$_2$O and CO$_2$ as competing mechanisms. The origin of this discrepancy is unclear.

Another type of competing reaction would be the destruction of methanol by photodissociation. The photofragments could then react to synthesise more complex molecules. In dense cloud conditions, photodissociation mainly occurs via cosmic ray induced photons, with a typical flux of $5\times 10^{3}$~photons~cm$^{-2}$~s$^{-1}$ \citep{Cecchi-Pestellini:1992}. Using a photodissociation cross section of $1.6\times 10^{-18}$~cm$^{2}$ \citep{Gerakines:1996, Oberg:methanol}, 5~\% of the methanol molecules has been photodissociated in $2 \times 10^{5}$~years. Most of these molecules are dissociated to simpler species like CO, HCO, CH$_{3}$, CH$_{3}$O, and H$_2$CO and can be hydrogenated to methanol again. Gas-phase H atoms can hydrogenate the radicals in top layers of the ice, whereas H atoms that are produced through photodissociation can react with the fragments deep in the ice. We therefore expect only a minor change in the grain abundances due to photodissociation as long as the grains remain cold.  Higher temperatures allow the radicals to diffuse rapidly enough to form large molecules such as methyl formate and dimethyl ether \citep{Garrod:2008, Oberg:methanol}. 

\section{Conclusions}
The surface formation of CH$_3$OH and H$_2$CO from precursor CO has been simulated using the continuous-time, random-walk Monte Carlo method. The formation of both species was found to be very efficient under certain conditions and to depend mainly on grain temperature and the gas phase abundance ratio of H and CO. During the freeze-out of CO onto the grain, this ratio changes, favouring the more complete hydrogenation of CO to CH$_3$OH. The more unsaturated species remain locked in the lower layers of the ice mantle. Due to the layering of the ice, changes in the gas phase abundance can only affect the top layers of the grain, which is important to take into account when modelling grain surface chemistry.   The detailed Monte Carlo method used can be compared with a variety of other approaches.  Of these, perhaps the most successful is a quasi-steady-state rate equation approach \citep{Charnley:1997} that focuses on the outermost layer of the grain only.

The model results can be compared with observations through plots of the CO/CH$_3$OH and H$_2$CO/CH$_3$OH abundance ratios. A very good agreement is obtained for the outer envelopes of a number of YSO's, with temperatures in the range 12.0-16.5~K. Moreover, the comparison allows us to trace the temporal evolution between different sources. Our results suggest that the difference in CH$_3$OH abundances with respect to water are mostly due to  differences in temporal evolution where the younger sources have not yet had sufficient time to build up much methanol.

\section*{Acknowledgements}
H.~C.~is supported by the Netherlands Organization for Scientific Research (NWO) and the Leiden Observatory. E.~H.~thanks the National Science Foundation (US) and NASA for support of his research programs in astrochemistry and astrobiology.  We would like to thank Lars Kristensen, Karin \"Oberg, and Harold Linnartz for stimulating discussions.

\begin{appendix}
\section{CO hydrogenation in steady state conditions}
This appendix presents a derivation of the steady state model proposed by \cite{Charnley:1997}.
When the flux of hydrogen atoms to the surface is lower than the flux of CO molecules, the assumption can be made that every hydrogenation atom either reacts with CO or with H$_2$CO. The probabilities of reaction are denoted by $\chi_{\rm CO}$ and $\chi_{\rm H_2CO}$, respectively.  They are determined by the likelihood of barrier crossing once an H atom is in the vicinity of the CO or H$_2$CO, $P_{\rm CO}$ and $P_{\rm H_2CO}$, and the fractional coverage of these species over the outermost surface layer, $\theta_{\rm CO}$ and $\theta_{\rm H_2CO}$ according to the equations
\begin{equation}
\chi_{\rm CO} = \frac{P_{\rm CO}\theta_{\rm CO}}{P_{\rm CO}\theta_{\rm CO} + P_{\rm H_2CO}\theta_{\rm H_2CO}} 
\label{Xco}
\end{equation}
and
\begin{equation}
\chi_{\rm H_2CO} = \frac{P_{\rm H_2CO}\theta_{\rm H_2CO}}{P_{\rm CO}\theta_{\rm CO} + P_{\rm H_2CO}\theta_{\rm H_2CO}}. 
\label{Xh2co}
\end{equation}
The change in CO coverage with time is then
\begin{equation}
\frac{{\rm d}\theta_{\rm CO}}{{\rm d}t} = f_{\rm CO}(1- \theta_{\rm CO}) - 2f_{\rm H}\chi_{\rm CO}.
\label{1}
\end{equation}
Here the first term accounts for the increase by the incoming CO flux, $f_{\rm CO}$ where the $- \theta_{\rm CO}$ accounts for covering of surface CO by new CO and the second, last term is due to reaction of CO with H. The factor two accounts for the two successive hydrogenation reactions to form H$_2$CO. In a similar fashion, rate equations for the surface coverage of H$_2$CO and CH$_3$OH can be determined to be
\begin{equation}
\frac{{\rm d}\theta_{\rm H_2CO}}{{\rm d}t} =  2f_{\rm H}(\chi_{\rm CO} - \chi_{\rm H_2CO}) - f_{\rm CO}\theta_{\rm H_2CO}
\label{2}
\end{equation}
and
\begin{equation}
\frac{{\rm d}\theta_{\rm CH_3OH}}{{\rm d}t} =  2f_{\rm H}\chi_{\rm H_2CO} - f_{\rm CO}\theta_{\rm CH_3OH}.
\label{3}
\end{equation}
From Eqs.~\ref{2} and \ref{3},  $\theta_{\rm H_2CO}/\theta_{\rm CH_3OH}$ is given by
\begin{equation}
\frac{\theta_{\rm H_2CO}}{\theta_{\rm CH_3OH}} =  \phi_{\rm H}\frac{\theta_{\rm CO}}{\theta_{\rm H_2CO}} - 1
\label{h2co/ch3oh}
\end{equation}
with
\begin{equation}
\phi_{\rm H} \equiv \frac{P_{\rm CO}}{P_{\rm H_2CO}} 
\end{equation}
and furthermore
\begin{equation}
\frac{\theta_{\rm CO}}{\theta_{\rm CH_3OH}} = \frac{\theta_{\rm H_2CO}}{\theta_{\rm CH_3OH}}  \frac{\theta_{\rm CO}}{\theta_{\rm H_2CO}}.
\label{co/ch3oh}
\end{equation}
From Eqs.~\ref{1} and \ref{2} and using Eqs.~\ref{Xco}-\ref{Xh2co}, we obtain that 
\begin{equation}
\frac{\theta_{\rm CO}}{\theta_{\rm H_2CO}} = \frac{1 + 2\phi_{\rm H}/\alpha_{\rm H} - \phi_{\rm H} + \sqrt{(1 + 2\phi_{\rm H}/\alpha_{\rm H} - \phi_{\rm H})^2 + 8\phi_{\rm H}/\alpha_{\rm H}}}{2\phi_{\rm H}}
\end{equation}
where
\begin{equation}
\alpha_{\rm H} \equiv \frac{f_{\rm H}}{f_{\rm CO}}.
\end{equation}
\end{appendix}

\end{document}